# Excitons in bulk black phosphorus evidenced by photoluminescence at low temperature

E. Carré[1,2], L. Sponza[1], A. Lusson[2], I. Stenger[2], E. Gaufrès[3], A. Loiseau[1], J. Barjon[2]

[1] LEM (Laboratoire d'Etude des Microstructures), ONERA-CNRS, UMR104, Université de Paris-Saclay, BP 72, 92322 Châtillon, France

[2] GEMaC (Groupe d'Etude de la Matière Condensée), CNRS-UVSQ, 45 Avenue des Etats-Unis 78035 Versailles, France

[3] LP2N (Laboratoire Photonique Numérique et Nanosciences), CNRS-Université de Bordeaux-Institut d'Optique, Rue François Mitterrand 33400 Talence, France

**ABSTRACT :**

Atomic layers of Black Phosphorus (BP) present unique opto-electronic properties dominated by a direct tunable bandgap in a wide spectral range from visible to mid-infrared. In this work, we investigate the infrared photoluminescence of BP single crystals at very low temperature. Near-band-edge recombinations are observed at 2 K, including dominant excitonic transitions at 0.276 eV and a weaker one at 0.278 eV. The free-exciton binding energy is calculated with an anisotropic Wannier-Mott model and found equal to 9.1 meV. On the contrary, the PL intensity quenching of the 0.276 eV peak at high temperature is found with a much smaller activation energy, attributed to the localization of free excitons on a shallow impurity. This analysis leads us to attribute respectively the 0.276 eV and 0.278 eV PL lines to bound excitons and free excitons in BP. As a result, the value of bulk BP bandgap is refined to 0.287 eV at 2K.

Since graphene was first isolated by mechanical exfoliation of graphite in 2004 [1], the 2D materials panorama has been considerably broadened. Black phosphorus (BP), a crystal first synthetized by Bridgman a hundred years ago [2], has recently been in the spotlight of the scientific community due to the unique properties of its 2D atomic layers: high carriers mobilities [3,4,5], in-plane anisotropy [6,7,8,9] and highly tunable bandgap [7,10,11]. In particular, the bandgap greatly depends on the BP thickness and goes from mid-infrared (IR) energy range in the bulk [12,13,14,15,16,17] to visible light energy range in the monolayer [4,5,10,18,19]. One can then observe that BP completes the accessible spectral range between the zero bandgap graphene [20,21] and semiconducting transition metal dichalcogenides (TMDCs) which exhibit luminescence in the visible range [22,23]. Furthermore and unlike 2D TMDCs [24,25], the BP bandgap remains direct whatever the layer number [26,27] which makes it interesting for optoelectronic applications such as photodetectors [28,29,30,31] or light-emitting devices [32,33,34]. BP exfoliated thinlayers have been intensively investigated by means of different optical measurements [12,35], electron energy loss spectroscopy [19] and especially through photoluminescence (PL) [4,5,36,37,38,39,40,41]. On the contrary, optical investigations on bulk BP are relatively old and quite sparse [42,43,44,45,46,47]. In particular, the characteristics of emitted light at low temperatures remains unknown, probably due to the technical difficulties related to the IR spectral region.

Here we report photoluminescence spectra at very low temperature of BP crystals keep as-synthetized. These measurements provide evidence of several contributions from near-band-edge recombinations, which are further investigated by temperature-dependent experiments. The highest energy PL peaks are attributed to the radiative recombination of excitons. An estimate of the binding energy of free excitons and their spatial extension has been done by means of a fully anisotropic Wannier-Mott model. The comparison between experimental and theoretical results is used as a basis for the characterization of the observed PL lines.

Bulk BP crystals were purchased from HQ Graphene (99.995% purity) and stored in glove box under argon atmosphere (<0.5 ppm $O_2$, <1 ppm $H_2O$) to prevent its photo-oxidation under ambient conditions. For luminescence experiments, PL is performed at 2 K thanks to a helium-bath cryostat at low pressure to reach the superfluid He state. The excitation source is a Nd:YAG (1064nm) laser focused on the BP crystal with a spot diameter of ~ 100 μm, well-suited for studying millimeter size BP crystals (see inset of Fig. 1). The penetration depth of the beam is of the order of 100 nm, defining a probed volume of $8.10^{-10}$ $cm^3$. The PL signal is collected in a Fourier Transform Infra-Red (FTIR) spectrometer (BOMEM DA8, $CaF_2$ beamsplitter, spectral resolution 0.5 meV) equipped with an InSb detector cooled at 80 K. A chopper and a lock-in amplifier are needed to eliminate the undesirable blackbody radiations in the infrared spectral range.

Figure 1 shows a low temperature (2 K) PL spectrum of a BP crystal. The intensity of the PL signal is of the same order of magnitude as that of a reference InAs crystal (see supporting information SI1). Note that the PL intensity depends on the polarization of the incident laser with respect to the zigzag and armchair axes of the BP atomic planes, due to the polarization-dependent absorption of BP crystals (See Fig. SI2) [7,8,9,10,11,12]. We also note that the sample is robust under beam exposure, no trace of degradation was observed after PL measurements.

The best fit of the PL spectrum at 2 K is obtained using four Gaussian contributions (see details in Fig. SI3): it reveals two narrow peaks at high energy (X at 0.278 eV and I°X at 0.275 eV) and two broader ones at lower energy (Y at 0.262 eV and Z at 0.251 eV). These values clearly correspond to near-band-edge recombinations in bulk BP, i.e. close to the bandgap values reported up to now (from 0.25 to 0.33 eV) [14,15,42,45,48,49]. Interestingly, the full-width at half-maximum (FWHM) of the higher energy peaks is particularly small (X : 1.7 meV and I°X : 2.9 meV). Moreover, their energy values are fully consistent with the 0.276 eV free-exciton energy extracted from reflectance experiments on bulk BP at 2 K [44,46]. These two elements allow us to attribute the higher energy peaks X (0.278 eV) and I°X (0.275 eV) to excitons of the bulk BP crystal. To the best of our knowledge, this is the first observation of excitons in bulk BP by photoluminescence. The two components I°X and X are less than 3 meV apart and might not be confused with the first two Rydberg series which are 6 meV apart [44].

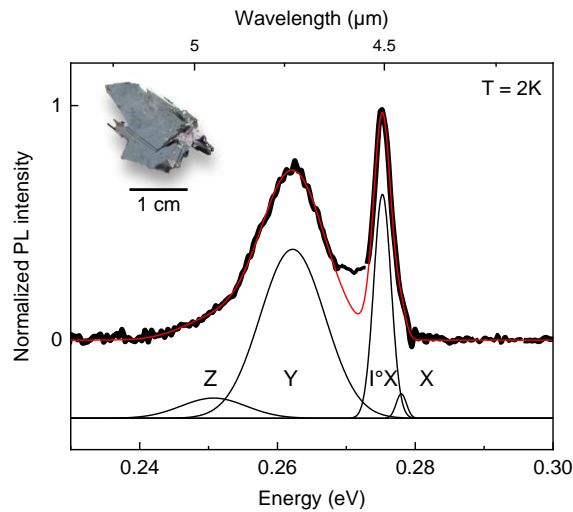

**Figure 1.** PL spectrum of bulk BP crystal at 2 K (dark line) excited by a 20 mW, 1064 nm laser. Data are fitted (red line) with four Gaussian curves to extract energies, linewidths and intensities (black lines shifted vertically for clarity). Inset: optical micrograph of the BP crystal.

Morita [44] evaluated the binding energy of free excitons in BP to be 7.9 meV when assuming BP to be an isotropic medium. However, BP is a highly anisotropic crystal with different effective masses [50] and dielectric constants [51], measured and reported along the three crystallographic axes related to the BP unit cell. In order to account for the 3D anisotropy of the crystal, we developed and solved a variational effective-mass model which generalizes the works by Baldereschi and Diaz [57] and Schindlmayr [58]. The Hamiltonian of the exciton is written as a sum of a kinetic part which accounts for the band effects (the anisotropic effective mass) and a potential energy accounting for screening effects (anisotropic dielectric function). We further assume that the excitonic wavefunction $\psi_a(r)$ has a 1s shape deformed as an ellipsoid along the three crystallographic axes. Hence, the exciton wavefunction depends on the parameter $a = (a_x, a_y, a_z)$ which quantify the axial deformation. Under these approximations, the excitonic ground state is found by minimizing the exciton energy $(a)=\langle\psi_a|H|\psi_a\rangle$ with respect to the parameter $a$. The value $\tilde{a}$ which minimizes $E(a)$ gives an estimate of the excitonic extension in the armchair (x), zigzag (y) and stacking (z) directions, and allows to evaluate the binding energy as $E_b = -E(\tilde{a})$. Refer to SI4 for additional information about the model.

On the basis of experimental input from literature on the dielectric function [51] and effective masses [50], the fully-anisotropic model predicts a free-exciton binding energy equal to 9.1 meV and excitonic extensions of 90.9 Å, 39.6 Å and 63.1 Å in the armchair (x), zigzag (y) and stacking (z) directions, respectively. Cuts on the *xy* and *xz* planes of the exciton wavefunction squared are reported in Fig. 2. It clearly illustrates the impact of the crystal anisotropy on the spatial extension of the exciton wavefunction. The average distance between the electron and the hole in the exciton complex appears strongly compressed along the zigzag axis. This is consistent with both the larger effective mass ($\mu_{\text{zigzag}} \simeq 10\mu_{\text{armchair}}$) and the lower dielectric constant ($\epsilon_{\text{zigzag}} = 13$, while $\epsilon_{\text{armchair}} = 16.5$). The extension along *z* (stacking axis) is also reduced with respect to the extension along the armchair axis, but not as much as in the zigzag direction. An anisotropy stronger within the plane than out of plane is a singular characteristic for a lamellar material, which points out once more how BP is rather unique among the 2D materials.

Additional simulations done using either an average isotropic dielectric constant or an average isotropic effective mass (see SI4) show that taking into account the anisotropy of the effective mass is crucial, while the anisotropy of the screening has a lower impact on the binding energy and the excitonic radii. This result is in contrast with what is generally observed in thin films, where image charges at the interfaces modify drastically the electron-hole interaction [59]. This indicates that there should be a cross-over thickness at which screening effects starts dominating over band effects. We expect this conclusion to be general, and not only related to BP.

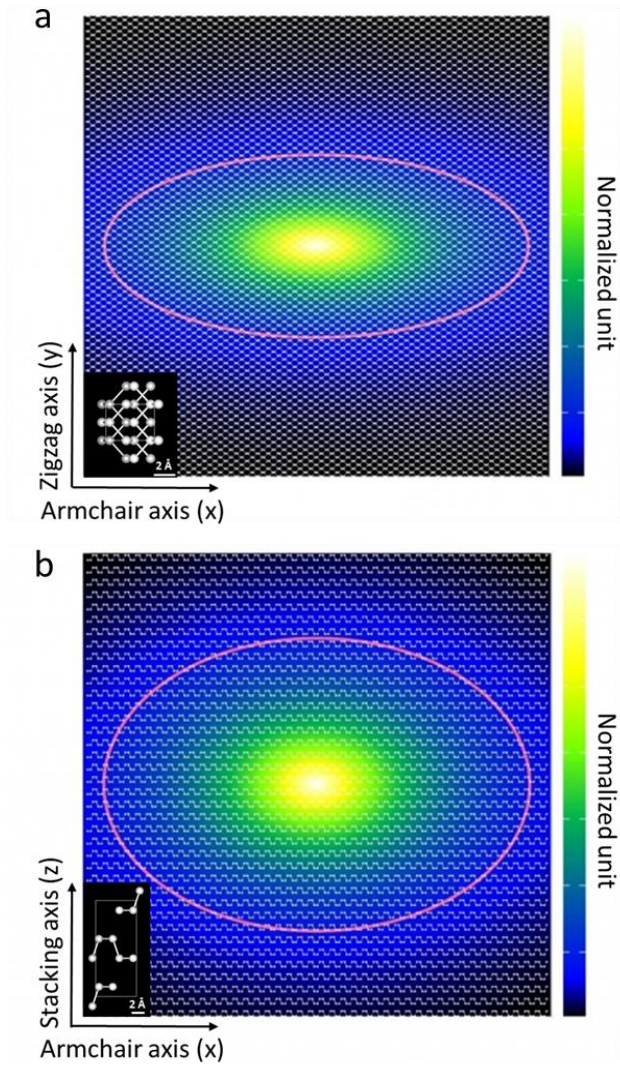

**Figure 2.** Planar cuts of the exciton wavefunction. (a) on the xy plane, (b) on the xz plane. Red ellipses are drawn from the optimized parameter $\tilde{a}$. Intensity units are arbitrary. The crystalline lattices in xy and xz planes are drawn in white in the insets.

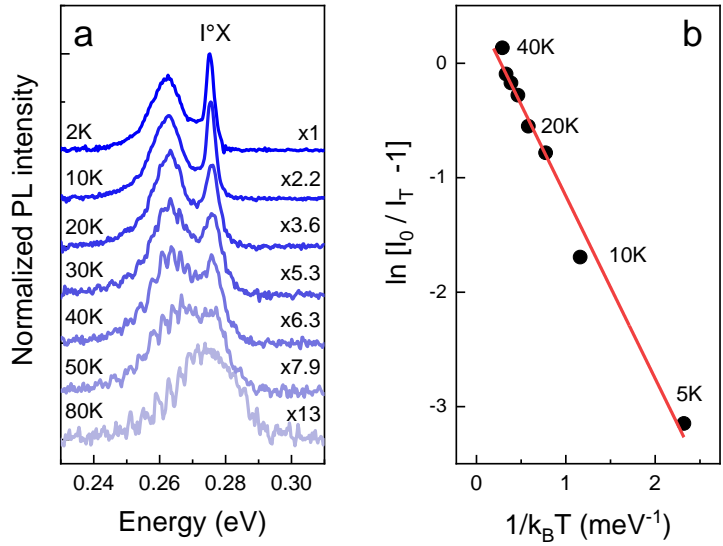

**Figure 3.** Temperature dependence of bulk BP photoluminescence. (a) PL spectra from 2 K up to 80 K. The decrease of I°X intensity at high temperature is modelled with a thermally-activated and non-radiative channel (see details in the text). A 1.6 meV activation energy is extracted from the linear fit (red line) in (b).

In order to investigate the exciton properties in more details, PL measurements as function of temperature were performed on another part of the crystal studied in Fig.1. These spectra are presented in Figure 3(a) from 2 K up to 80 K. The measurement at 2 K well reproduces the spectrum shown in Fig.1, attesting the reproducibility of the PL spectrum. At high temperature, the I°X and Y PL peaks shift towards higher energies, following the expected behavior of the BP bandgap (see also PL spectra up to 200 K and complementary analysis in SI5). The blue shift of the BP bandgap at high temperature is a well understood phenomenon [52], though a redshift is more commonly observed in semiconductors [44,48].

Besides, the I°X peak is distinguishable only up to 50 K and vanishes at higher temperatures. Its integrated intensity is analyzed in Figure 3(b). The I°X PL intensity is fitted with a simple model assuming a thermal activation of a non-radiative processes, $I(T)/I(0) = \left(1 + a \exp(-E_a/k_B T)\right)^{-1}$, where $a$ is the ratio between radiative and non-radiative recombination probabilities and $E_a$ is the activation energy of non-radiative channels [53,54,55]. In Fig. 3(b), $I(0)$ is assumed as the intensity measured at 2 K and the experimental data are plotted as $\ln(I(0)/(I(T)) - 1)$ as a function of reverse temperature. This plot evidences that a single activation energy is enough to account correctly for the I°X thermal quenching. A linear regression gives $E_a$ = 1.6 meV with a non-radiative parameter $a$

= 1.5. This activation energy is more than 5 times lower than the free exciton binding energy (9.1 meV), which rules out free-exciton recombinations as the origin of the I°X peak.

It is important to remind now that the purest BP crystals are naturally p-type doped, as demonstrated by the transport studies performed on bulk BP in the 1990s [14,48]. The nature of the acceptor center responsible for the non-intentional p-type conductivity of BP remains unclear but could be related to the presence of intrinsic vacancies [60]. Its ionization energy was found to be about 18 meV [48]. Interestingly, the localization energy of free excitons on shallow dopants is known to be typically 10% of the dopant ionization energy according to Haynes rule [56]. The activation energy extracted from the temperature-dependent experiments would rather correspond here to the localization energy of the free exciton X on a neutral impurity I°, to form a bound exciton complex I°+X –> I°X. For that reason, we attribute the I°X luminescence to the radiative recombination of the exciton bound to a shallow impurity. The shallow dopant is probably a neutral acceptor in the investigated sample, which deserves to be confirmed by Hall effect measurements. The temperature-dependence of the bound exciton linewidth was further analyzed and a coupling constant with acoustic phonons $\Gamma_a = 102$ μeV/K was found, with the same order of magnitude than in InP or CdSe (see SI5c).

The results obtained on the activation energy of the bound exciton PL quenching are consistent with the presence of a faint shoulder observed on the high-energy side of I°X peak in Fig.1. The careful deconvolution presented in Fig. 1 reveals a peak (the X peak) located at $2.5 \pm 1$ meV above the I°X peak, consistent with the order of magnitude of the localization energy extracted from temperature-dependent PL experiments. Though weak, the energy of the X PL signal at 278 meV might be associated to the free exciton of BP.

Lastly, we briefly discuss the origin of the lower energy PL contributions, Y and Z, observed respectively at 16 and 27 meV lower than the bound exciton I°X. The dominant contribution Y might correspond to a transition from electron to neutral acceptor (eA$^0$), or donor-acceptor pairs (DAP) recombinations since a 18 meV acceptor was identified in non-intentionally doped BP crystals [14,48]. The lower PL energy contribution, Z, could come from DAP transitions. Despite experimental evidences consistent with these hypotheses showing that Y and Z peaks vary both in energy and in intensity as function of the excitation power (see S6-7), their complete attributions would deserve further experimental work.

To conclude, our work shows the first PL spectrum of black phosphorus crystal at very low temperature and, in contrast to exfoliated bulk layers, the PL spectrum is composed of several peaks between 4.5 and 5 μm. The experimental and theoretical results gathered in this work confirm the excitonic nature of the dominant I°X peak, as a bound exciton with a localization energy of 2.5 meV. This result is reinforced by the fact that it is obtained using two strictly independent methods, which are the activation energy of peak I°X quenching and the observation of peak X at this same energy. These experimental results associated with the calculation of the free-exciton binding energy provide a revised and more accurate value of $0.287 \pm 0.001$ eV for the free particle bandgap of bulk BP in the limit of low temperatures. This value deserves to be the bulk reference value for further investigations of thickness dependent properties in thin films and layers.

The authors acknowledge funding from the French national research agency (ANR) under the grant agreement No. ANR-17-CE24-0023-01 (EPOS-BP). This project has also received funding from the European Union's Horizon 2020 research and innovation program under grand agreement N° 785219 (Graphene Flagship core 2) and N° 881603 (Graphene Flagship core 3).

# Supporting information

## 1) Photoluminescence of bulk Black Phosphorus compared to InAs

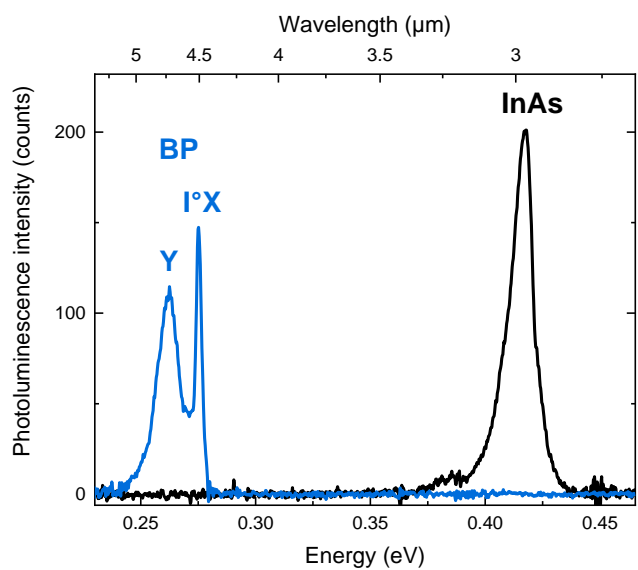

**Figure SI1.** PL spectra of the bulk BP crystal (c.f. Fig. 1 of the main manuscript) compared to a reference InAs crystal measured in the same conditions at 2 K, with a 20 mW, 1064 nm laser.

Figure SI1 shows low temperature PL spectra of BP and InAs single crystals. The two samples were placed on the same holder and measured under the same conditions (2 K, 20 mW, 1064 nm laser). The PL spectrum of InAs shows a narrow peak at 0.417eV and a weak shoulder at lower energy near 0.38eV. The InAs luminescence features are well-known from previous PL reports [1,2]. The narrow peak is attributed to excitonic recombination while the low-energy shoulder is due to deep impurities or defect related acceptor levels [3]. Importantly, the BP crystal shows a total integrated PL intensity only 50% lower than the one of the InAs reference, which attests the good quality of the BP crystal investigated in this work.

## 2) Effect of the laser polarization on the PL of Black Phosphorus

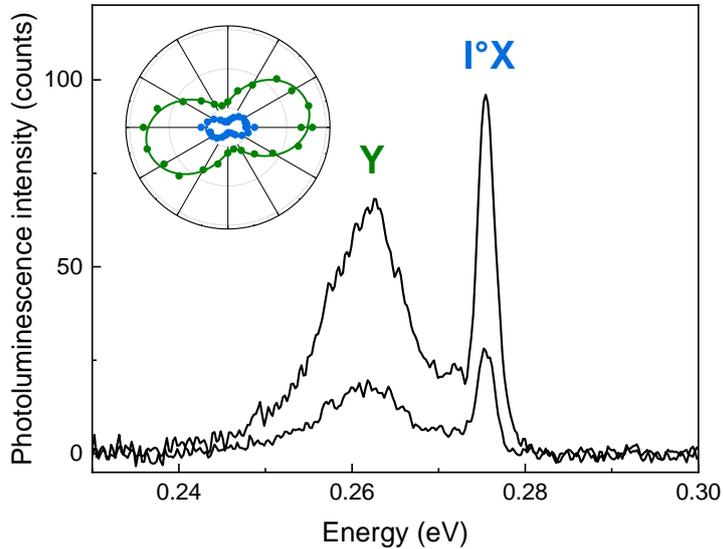

**Figure SI2.** Dependence of BP PL at 2K toward the linear polarization of the laser. The reported PL spectra are those corresponding to the incident polarizations giving rise to maximum and minimum intensities. The angular diagram of integrated intensities of $I^0X$ (blue) and Y (green) lines is plotted in inset as a function of the polarization angle of the laser in the sample plane.

The results plotted in Figure SI2 were obtained with the introduction of a λ/4 plate after the laser to have a circular polarization. A rotating polarizer was placed after it to vary the angle of the linear incident polarization on the sample. It is found that the ratio I°X / Y remains constant whatever is the incident polarization, while the total PL intensity varies within a factor ~ 3. This effect is probably due to the in-plane anisotropy of BP, for which the absorption coefficient is polarization-dependent as already stated in Refs [4,5]. To preserve the excitation power from a strong absorption in polarization plates, in the main body of the manuscript the experiments have been performed without controlling the direction of the laser polarization on the BP crystals.

## 3) Fitting procedures

The fitting of the PL spectrum in Fig. 1 is achieved by a least-squares minimization with free parameters (energy, linewidth, area) and assuming the two low energy peaks share the same linewidth. Note that energy ranges have been excluded from the data analysis: on 0.268-0.272 eV an artefact signal is identified and 0.288-0.295 eV is the $CO_2$ absorption range. The best fit of the PL spectrum at 2K is found when using 4 Gaussian curves, giving a R-square coefficient as high as 99.79 %. The results are summarized in the table below:

| **Energy** (eV) | 0.250(7) | 0.262(2) | 0.275(2) | 0.277(9) |
|---|---|---|---|---|
| **FWHM** (meV) | 10.8 | 10.8 | 2.9 | 1.7 |
| **Area** | 0.36 | 1 | 0.12 | 0.02 |

The weak component at 0.277(9) eV in Fig. 1 might be hard to distinguish at first glance. To better evidence it, the PL spectrum is fitted with only 3 gaussians in Fig. SI3. A deviation appears at high energy (c.f. black arrow). It is consistent with the R-square value of 97.90 % that is smaller than for the previous 4-gaussian fit. The residue from fitted curves is plotted in inset of Fig. SI3 to better isolate the higher-energy PL contribution.

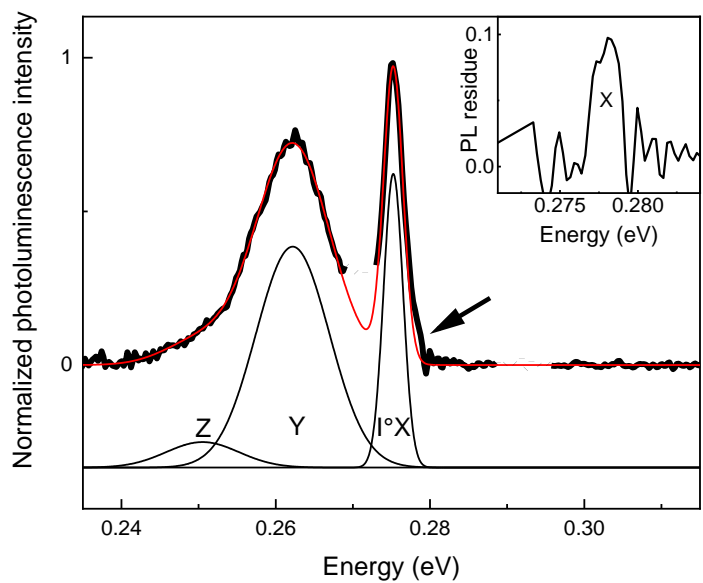

**Figure SI3.** Fitting the PL data of Fig. 1 with only 3 Gaussians evidences a deviation at high energy (black arrow). Inset : plot of the fit residue to isolate the high energy contribution to the PL spectrum (X).

Due to the lower signal-to-noise ratio in temperature-dependent experiments, the minimization of chi-square is not converged when using four gaussians. For the analysis of Fig. 3(b) and Fig. SI5, only 3 Gaussians are used.

## 4) Variational anisotropic effective-mass model

We treat the excitonic problem as an anisotropic hydrogenoid system. The 3D anisotropic Hamiltonian $H = E_\text{k} + V$ of the exciton reads

$$H(\boldsymbol{r}) = -\frac{\hbar}{2}\left(\frac{\partial^2}{\mu_x \partial x^2} + \frac{\partial^2}{\mu_y \partial y^2} + \frac{\partial^2}{\mu_z \partial z^2}\right) - \frac{e^2}{4\pi\epsilon_0 \sqrt{\epsilon_{yy}\epsilon_{zz}x^2 + \epsilon_{xx}\epsilon_{zz}y^2 + \epsilon_{xx}\epsilon_{yy}z^2}},$$

where $\mu_j$ is the reduced mass of the exciton along the direction $j$, $e$ is the electron charge, $(4\pi\epsilon_0)^{-1}$ is the Coulomb constant, and the biaxial dielectric tensor of the crystal $\epsilon_{ij}$ is diagonal on the Cartesian axes. Similarly to what done in other similar works [6,7,8] we further assume a 1s-like shape for the exciton wavefunction, namely

$$\psi_a(\boldsymbol{r}) = \sqrt{\frac{1}{\pi a_x a_y a_z}} \exp\left[-\sqrt{\left(\frac{x}{a_x}\right)^2 + \left(\frac{y}{a_y}\right)^2 + \left(\frac{z}{a_z}\right)^2}\right]$$

where $a_j$ is the exciton radius along the direction $j$. The radius parameter $\boldsymbol{a}$ is a variational parameter with respect to which the exciton energy $E(a_x, a_y, a_z) = \langle\psi_a|H|\psi_a\rangle$ has to be minimized. At the minimum energy $\tilde{E} = E(\tilde{\boldsymbol{a}})$, the excitonic extension on the Cartesian directions is given by the radius $\tilde{\boldsymbol{a}} = (\tilde{a}_x, \tilde{a}_y, \tilde{a}_z)$ and the corresponding binding energy is simply $E_b = -\tilde{E}$.

The kinetic energy term can actually be computed analytically in spherical coordinates, but the potential energy has to be minimized numerically. We do this by evaluating the integral in real space of $\langle\psi_a|V|\psi_a\rangle$ on a spherical grid of radius $R$ which is sampled densely close to the origin and more coarsely close to $R$. The number of points, the value of $R$ and the density of the grid are parameters to be converged.

Since the integrand function diverges at the origin, the contribution of the small volume $\Omega_0$ enclosing the origin has been approximated with the isotropic potential integrated inside a sphere of volume $\Omega_0$. In a similar way, the contributions to the integrals beyond the spherical grid ($r$>$R$) have been approximated with an isotropic potential.

The energy minimization is done by changing iteratively the parameter $\boldsymbol{a}$ and following $-\nabla_{\boldsymbol{a}} E(\boldsymbol{a})$. This is computed numerically at each step by finite differences and the line search method used to adjust the iterative step is based on the Barzilai-Borwein algorithm [9]. The minimization has been stopped after 1000 iterations which are largerly enough for all values to be converged.

The input parameters used for the simulation are summarized in the Table SI4 and come from references [10] and [11]. In order to gain a deeper inside in the characteristics of the exciton, two simulations have been done in addition to the fully anisotropic one. In the first one, only the effective mass is anisotropic, while the elements of the dielectric tensor are set to their average value $\epsilon_A = 12.6$. The second simulation keeps an anisotropic dielectric constant, but fixes the reduced effective mass of each component to its average value $\mu_A = 0.175\ m_0$. The comparison between these two calculations and the fully anisotropic one shows that the anisotropy of the effective mass determines the characteristics of the exciton much more than the anisotropy of the dielectric function.

| Input parameters | | | |
|---|---|---|---|
| Exciton masses ($m_0$) | 0.040 | 0.397 | 0.087 |
| Dielectric function | 16.5 | 13.0 | 8.3 |
| Volume at origin $\Omega_0$ (Bohr) | 0.25 | 0.25 | 0.25 |
| No. points in the axis | 300 | 300 | 300 |
| Radius of the grid R (Bohr) | 500 | | |

| Results | | | | |
|---|---|---|---|---|
| Anisotropy | $a_{armch.}$ (Å) | $a_{zigzag}$ (Å) | $a_{stack}$ (Å) | $E_b$ (meV) |
| Fully anisotropic | 90.9 | 39.6 | 63.1 | 9.1 |
| $\epsilon_A = 12.6$ | 91.1 | 40.7 | 68.7 | 8.3 |
| $\mu_A = 0.175\ m_0$ | 38.6 | 37.2 | 35.0 | 16.0 |

## 5) Temperature dependence of bulk BP photoluminescence

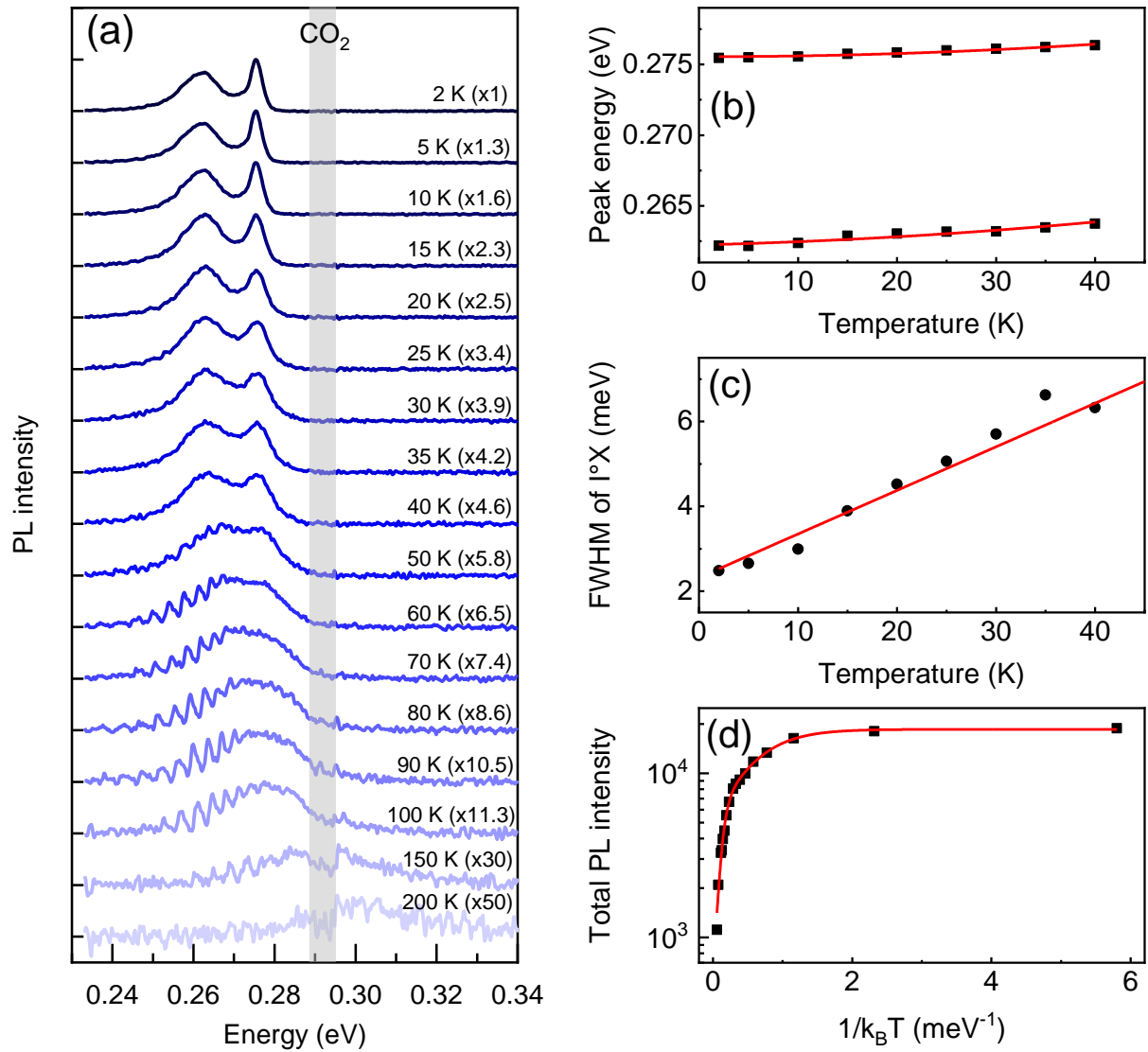

**Figure SI5.** (a) Temperature dependence of the PL spectrum recorded from the bulk BP crystal between 2 K and 200 K under a 20 mW, 1064 nm laser excitation. The normalization factors are indicated in parenthesis. (b) PL energy of I°X and Y peaks as a function of temperature fitted with the Varshni model (details in the text). (c) FWHM of the I°X peak as a function of the temperature, the red curve is a fit accounting for the broadening by exciton-acoustic phonon interactions (details in the text). (d) Arhenius plot of the total integrated PL intensity as a function of inverse temperature, fitted (red curve) with the thermal activation of two non-radiative processes (details in the text).

Figure SI5(a) shows the PL spectra at different temperatures ranging from 2 K to 200 K. The decrease of PL intensity at high temperature by almost 2 orders of magnitude, revealing the activation of non-radiative processes overcoming PL emission. An high energy shift occurs for I°X and Y peaks, which is more clearly observed for Y over a wider temperature range. The PL intensity drop, the energy shifts and the broadening of the exciton linewidth at high temperature are analyzed in the following. Oscillations are observed in the low-energy part of PL spectra from 50K upwards, which may be due to interference phenomena.

Figure SI5(b) shows the evolution of the I°X and Y peak PL energy with temperature from 2 K to 40 K. Both peaks show a shift towards higher energies as the temperature rises. The shift reflects the temperature dependence of the bandgap energy as shown in previous reports [12,13]. Up to know it has been only analyzed with linear fits over limited temperature ranges. Instead here, the I°X and Y peak energies are simultaneously fitted with the classic and empirical Varshni model, well-known in semiconductor physics :

$$E_i(T) = E_i(0) - \alpha \frac{T^2}{T+\beta},$$

where $E_i(0)$ is assumed as the peak energy observed at 2 K, T the temperature in Kelvin, α and β are the fitting parameters. As Varshni indicated in his original paper [14], the parameter b is usually close to the Debye temperature of the crystal, estimated around 300 K at low temperature for BP [15,16]. On the 2-40 K range, T is thus negligible compared to beta and the Varshni model can then be written : $E_i(T) = E_i(0) - \frac{\alpha}{\beta}.T^2$. The energy dependence of I°X and Y are both well-fitted with $\frac{\alpha}{\beta} = -0.6\ \mu eV/K^{-2}$. This indicates that the Y peak energy follows BP bandgap, as the I°X exciton peak. Assuming beta equal to 300 K, the order of magnitude of alpha is $10^{-4}$ eV/K, which is the typical order of magnitude for small bandgap semiconductors. Note that, the alpha value is found negative for BP. This is a quite rare situation in semiconductors (for instance in InAs α=+ 3.16.$10^{-4}$ eV/K [14]). This particular behavior of BP is well understood theoretically [17].

Figure SI5(c) shows the broadening of the excitonic peak I°X as a function of the temperature from 2 K to 40 K. It is usually analyzed with the following expression for exciton recombinations in direct bandgap semiconductors :

$$\Gamma(T) = \Gamma(0) + \Gamma_a.T + \frac{\Gamma_{LO}}{e^{\frac{E_{LO}}{k_B.T}}-1}$$

where $\Gamma$ is the full width at half maximum (FWHM) of the exitonic peak, $\Gamma(0)$ is the temperature-independent inhomogeneous broadening, the term $\Gamma_a$ is the exciton-acoustic phonon coupling constant and the term $\Gamma_{LO}$ represents the exciton-LO phonons coupling coefficient where $E_{LO}$ is an average energy for LO phonons. In BP $E_{LO}$ is in the range of 45 meV so that the coupling of excitons to acoustic phonons has a negligible contribution in the 2-40 K temperature range [18]. A linear temperature dependence describes well indeed the experimental data with $\Gamma_a = 102$ µeV/K. This value has the typical order of magnitude observed for other semiconductors like InP ($\Gamma_a = 400$ µeV/K) or CdSe ($\Gamma_a = 60$ µeV/K) [19].

Figure S5(d) shows the total PL intensity quenching at temperature as high as 200 K. The PL intensity decreases with increasing temperature due to a competition with non-radiative recombination channels. The decay of the total PL intensity is fitted here with a Gurioli model [20] with two activation energies : $I(T)/I(0) = \left(1 + a_1 \exp\left(-\frac{E_{a1}}{k_B T}\right) + a_2 \exp\left(-\frac{E_{a2}}{k_B T}\right)\right)^{-1}$, where $a_i$ are the ratios between radiative and non-radiative recombination probabilities and $E_{ai}$ are the activation energies of non-radiative channels. The best fit is found with $E_{a1}$=2.4 meV, $E_{a2}$=19.4 meV and $a_1$=2.5, $a_2$=30.4. We note that the low activation energy is the same order of magnitude than the localization energy of the bound exciton I°X, which is consistent with the analysis on Fig. 3 in the main manuscript. The higher energy, responsible for the decrease of the total PL signal at higher temperature, is close to the ionization energy of the acceptor defects measured by Baba and Maruyama (18meV) [21,22], which highlights again the probable role of shallow dopants at the origin of Y and Z PL peaks. We note that the higher energy is close to the activation energy recently found by PL analysis of thick BP layers by Zhang [23].

## 6) Power dependence of bulk BP photoluminescence

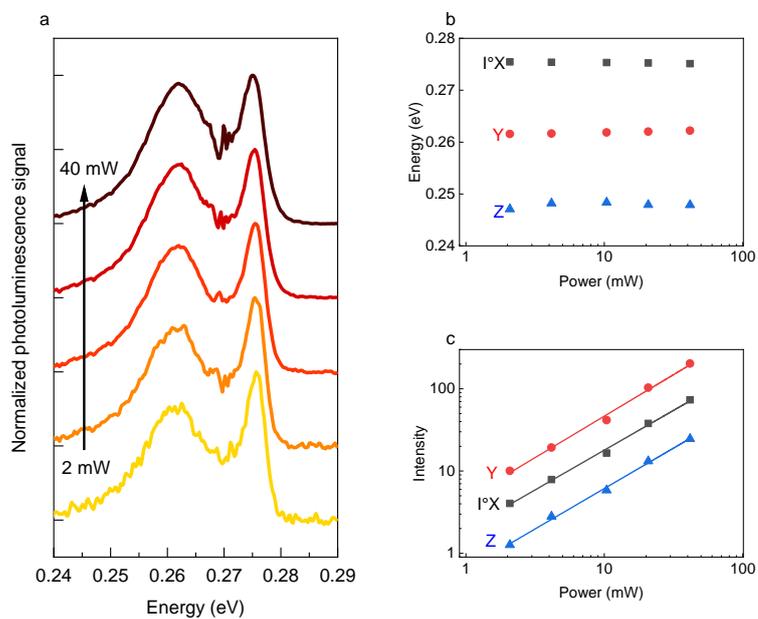

**Figure SI6.** (a) Power dependence of the PL at 2K measured from another BP crystal than the one studied in the main manuscript. (b) PL energy and (c) PL intensity of I°X, Y and Z peaks as a function of excitation power.

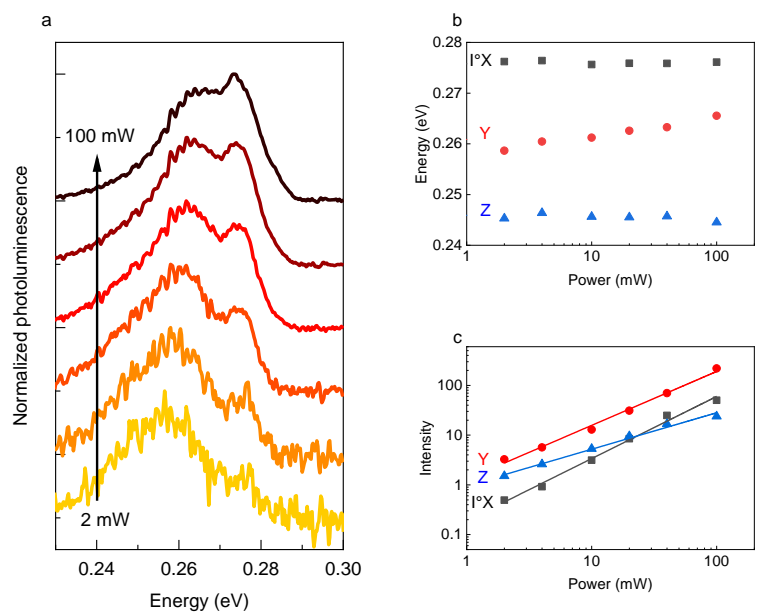

**Figure SI7.** (a) Power dependence of the PL at 2K from the same crystal as in the main manuscript, but degraded by non intentional air exposition. (b) PL energy and (c) PL intensity of I°X, Y and Z peaks as a function of excitation power.

The two figures SI6(a) and SI7(a) show PL spectra at 2K on BP crystals studied as a function of the incident laser power measured on the samples. The measurements of Figure SI6(a) were performed on a different crystal than the one presented in the main manuscript, but from the same supplier. The PL spectra turn out to be almost identical with a bound exciton peak at 0.275 eV and lower energy components at 0.248 and 0.262 eV. The measurements of Figure SI7(a) were carried out on the same crystal than the one presented in the main manuscript, but 6 months later. The global PL intensity appear much weaker, especially its I°X component. This is probably related to a degradation of the BP crystal despite the care taken for stocking it, as reported by many authors [24,25].

The analysis of peak energies and intensities is reported in (b) and (c) of the corresponding figures after the deconvolution procedure described in part SI3. The invariance in energy of the bound exciton peak (I°X) towards the incident power indicates that the heating of the sample is not significant in the power range investigated. A blue shift of low energy PL components when increasing the incident power is observed in Figure SI7(b)).This behavior generally points out a donor-acceptor pair (DAP) transition.

The PL intensities are fitted with a classical linear model $I = P^k$ with I is the integrated intensity of the PL peaks, P the incident laser power and k a factor depending on the nature of the radiative transition [26,27]. For the I°X peak, we found k= 0.98 in Fig. SI6 and 1.35 in Fig. SI7, which is in agreement with the expected values for excitonic recombinations. For the Y peak, we found k= 1.00 (Fig. SI6) and 1.14 (Fig. SI7) which could be consistent with an electron to neutral acceptor (eA°) transition. Finally, for the Z peak, k is below 1 (0.98 for Fig. SI6 and 0.76 for Fig. SI7) as expected for donor-acceptor pair transitions.

In spite of contrasted results from the two samples, surely due to different crystal qualities, both measurements seem to agree on the fact that Y and Z are near band edge recombinations that could be defect-related, such as eA° or DAP transition. Further study would be required to clarify the nature of the low-energy peaks and the effect of crystal degradation on them.